\documentclass[letterpaper]{jpsj-suppl}
\usepackage{txfonts} %Please comment out this line unless the txfonts package is availabe in your LaTeX system.
\usepackage{amssymb}   % for math
\usepackage{amsmath}
\pdfoutput=1

\title{Low-Background In-Trap Decay Spectroscopy with TITAN at TRIUMF}

\author{K.G.~Leach$^{1,2}$, A.~Lennarz$^{1,3}$, A.~Grossheim$^1$, R.~Klawitter$^{1,4}$, T.~Brunner$^{1,5}$, A.~Chaudhuri$^1$, U.~Chowdhury$^{1,6}$, J.R.~Crespo~L\'opez-Urrutia$^4$, A.T.~Gallant$^{1,7}$, A.A.~Kwiatkowski$^1$, T.D.~Macdonald$^{1,7}$, B.E.~Schultz$^1$, S.~Seeraji$^2$, C.~Andreoiu$^2$, D.~Frekers$^3$, and J.~Dilling$^1$}

\inst{
  $^1$TRIUMF, 4004 Wesbrook Mall, Vancouver, BC, V6T 2A3, Canada\\
  $^2$Department of Chemistry, Simon Fraser University, Burnaby, BC, V5A 1S6, Canada\\
  $^3$Institut f\"ur Kernphysik, Westf\"alische Wilhelms-Universit\"at, D-48149 M\"unster, Germany\\
  $^4$Max-Planck-Institut f\"ur Kernphysik, D-69117 Heidelberg, Germany\\
  $^5$Department of Physics, Stanford University, Stanford, CA, 94305, USA\\
  $^6$Department of Physics and Astronomy, University of Manitoba, Winnipeg, MB, R3T 2N2, Canada\\
  $^7$Department of Physics and Astronomy, University of British Columbia, Vancouver, BC, V6T 1Z1, Canada
}

\email{kleach@triumf.ca}

\recdate{\today}

\abst{An in-trap decay spectroscopy setup has been developed and constructed for use with the TITAN facility at TRIUMF.  The goal of this device is to observe weak electron-capture (EC) branching ratios for the odd-odd intermediate nuclei in the $\beta\beta$ decay process.  This apparatus consists of an up-to 6~Tesla, open-access spectroscopy ion-trap, surrounded radially by up to 7 planar Si(Li) detectors which are separated from the trap by thin Be windows.  This configuration provides a significant increase in sensitivity for the detection of low-energy photons by providing backing-free ion storage and eliminating charged-particle-induced backgrounds.  An intense electron beam is also employed to increase the charge-states of the trapped ions, thus providing storage times on the order of minutes, allowing for decay-spectroscopy measurements.  The technique of multiple ion-bunch stacking was also recently demonstrated, which further extends the measurement possibilities of this apparatus.  The current status of the facility and initial results from a $^{116}$In measurement are presented.}

\kword{in-trap decay spectroscopy, electron-beam ion trap, multiple ion-bunch stacking, beta-decay of highly-charged-ions}

\begin{document}
\maketitle

\section{Introduction}
\subsection{$\beta\beta$ Decay Nuclear Matrix Elements (NMEs)}
Recent evidence of massive neutrinos has generated great interest in exotic nuclear decay modes~\cite{Cir13,Avi08}.  As a part of these studies, searches for the $0\nu$ mode of $\beta\beta$ decay are particularly interesting~\cite{Avi08}, since an observation of this mode would establish the neutrino as a Majorana particle.  If this decay mode is observed, the effective Majorana mass of the neutrino can be extracted if the nuclear matrix element (NME) that connects the initial and final $0^+$ states is known~\cite{Bar12}.  The calculation of $\beta\beta$ decay NMEs is the focus of current theoretical efforts and includes several different model descriptions.  These calculations can be constrained from experimental data such as measurements of the $\beta^-$ and electron-capture (EC) branching ratios of the intermediate nuclei in the $2\nu\beta\beta$-decay process.  Typically, EC transitions are several orders of magnitude weaker than the dominant $\beta^-$ decays from the same parent nucleus, making them difficult to measure.

In EC decay, the characteristic X-ray originates from the filling of the vacated atomic $K$-shell electron, and typically has an energy less than 100~keV.  For the cases of interest to $\beta\beta$-decay studies~\cite{Fre07} the X-rays are much lower in energy, and are generally less than 40~keV.  To observe weak EC branches at these energies, it is important that effects such as positron-annihilation, Compton, and charged-particle induced backgrounds are minimized.  Reducing photon backgrounds at these energies requires a high level of control over the decay environment which can be provided using ion traps.  Therefore a low-background, high-sensitivity decay-spectroscopy tool has been developed using the TITAN ion traps for measuring characteristic X-rays from weak EC decays~\cite{Fre07,Lea14}.

\section{TITAN at TRIUMF-ISAC}
\begin{figure}[t]
\begin{center}\includegraphics[width=\linewidth]{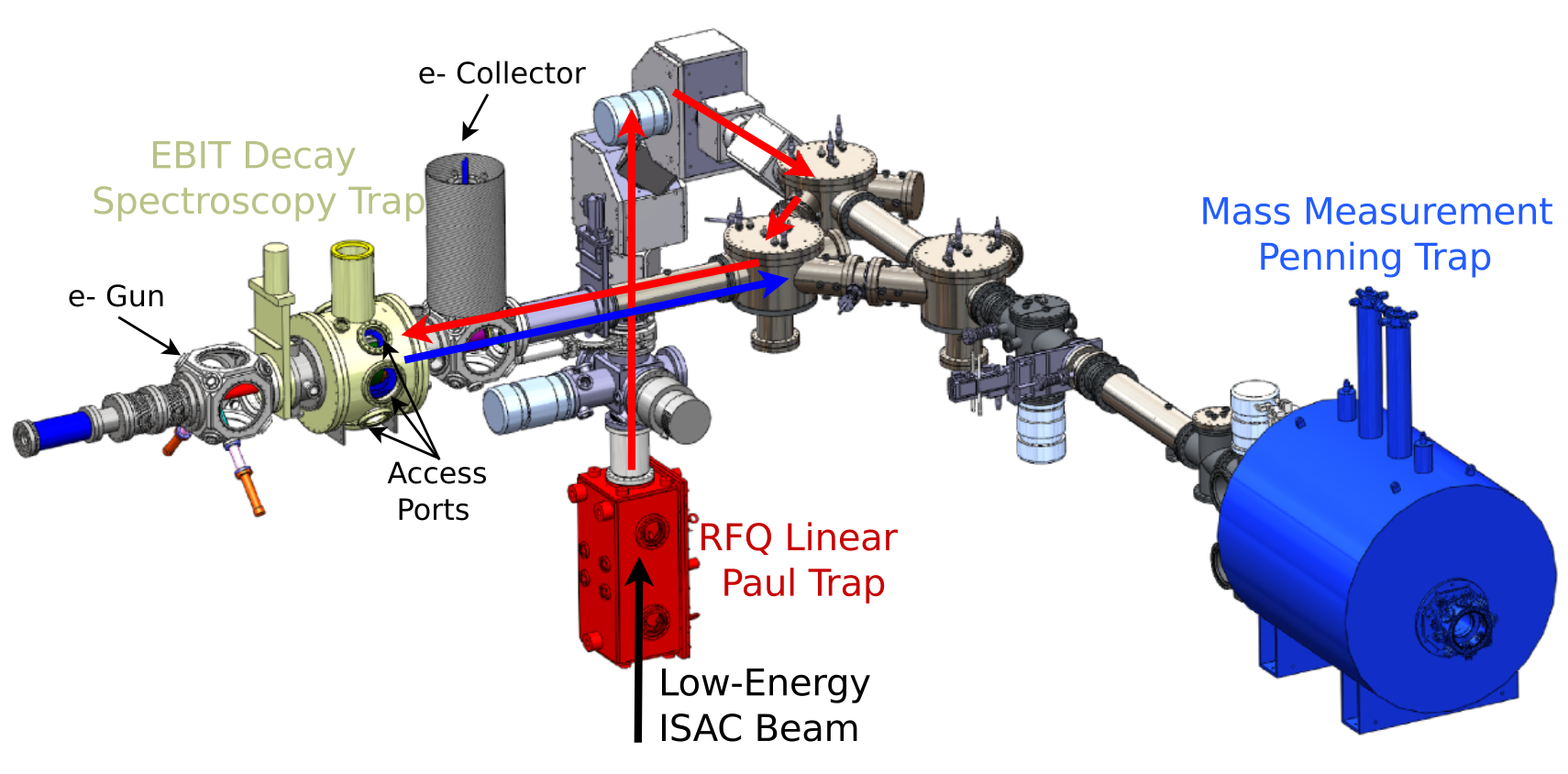}\end{center}
\caption{\label{TITAN}A schematic view of the TITAN facility at TRIUMF.  The ions are extracted in a bunch from the RFQ and injected as singly charged ions to the EBIT (red path) where they are stored for charge breeding and decay spectroscopy.  After the ions have been stored in the EBIT, they are extracted and dumped downstream away from the detectors (blue path).}
\end{figure}
TRIUMF's Ion Trap for Atomic and Nuclear Science (TITAN)~\cite{Dil06} consists of three ion traps; (i) an RFQ linear Paul trap~\cite{Bru12} for buffer-gas cooling and bunching the low-energy ion beam, (ii) a 3.7~T, high-precision mass-measurement Penning-trap (MPET)~\cite{Bro12}, and (iii) an electron-beam ion trap (EBIT) which provides highly charged ions (HCIs)~\cite{Lap10}.  TITAN resides at the Isotope Separator and Accelerator (ISAC) facility at TRIUMF, which employs a high-intensity (up to 100 $\mu$A) beam of 500 MeV protons to produce RIBs using the isotope separation on-line (ISOL) technique~\cite{Blu13}.  The mass-selected, continuous beam of radioactive singly charged ions (SCIs) is delivered at low energies ($<60$~keV) to a suite of experimental facilities for both cooled- and stopped-beam experiments~\cite{Dil14}.  The TITAN facility is primarily used to perform high-precision Penning-trap mass spectrometry on short-lived radioactive nuclides~\cite{Gal14,Fre13,Gal12,Bro12}, however this article describes a decay-spectroscopy setup using the TITAN ion traps (Fig.~\ref{TITAN}), and described in detail below.

\subsection{Decay Spectroscopy with TITAN}
\begin{figure}[t!]
\begin{center}
\includegraphics[width=0.405\linewidth]{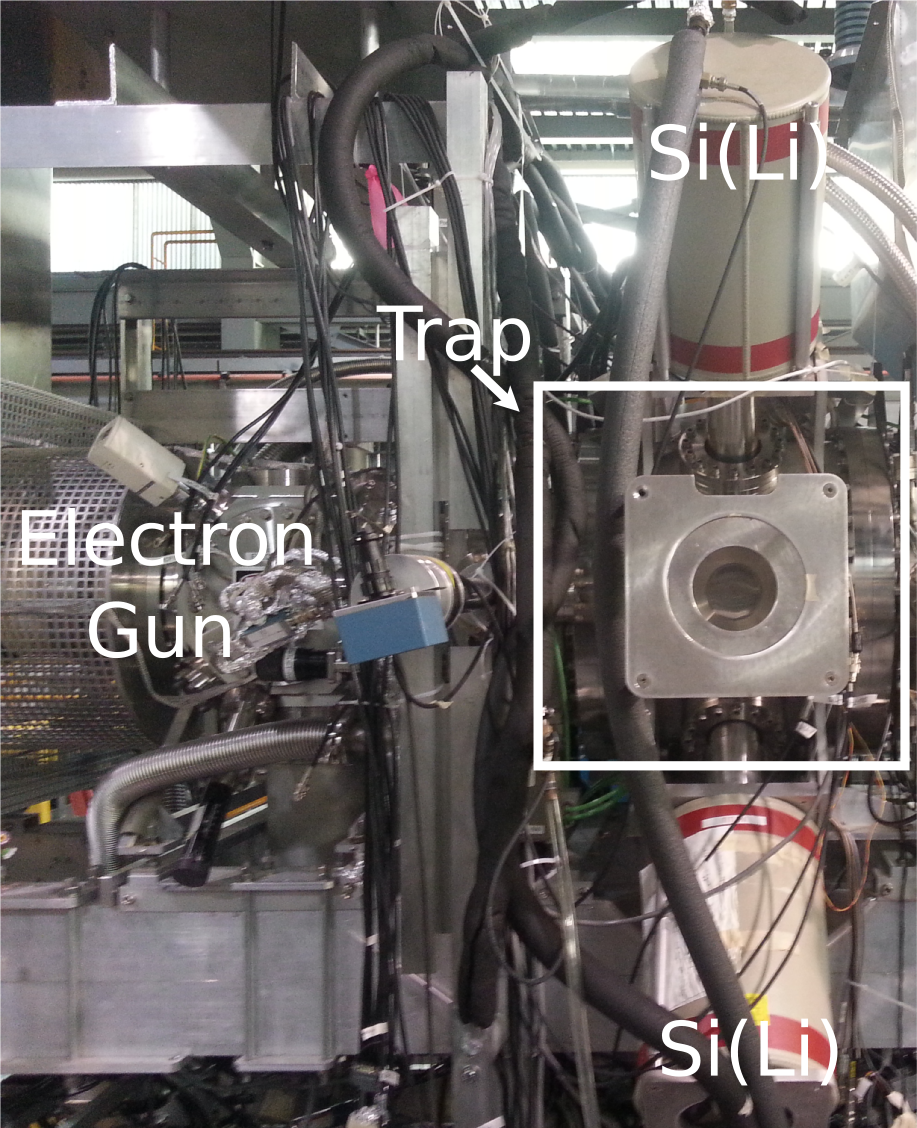}~\includegraphics[width=0.55\linewidth]{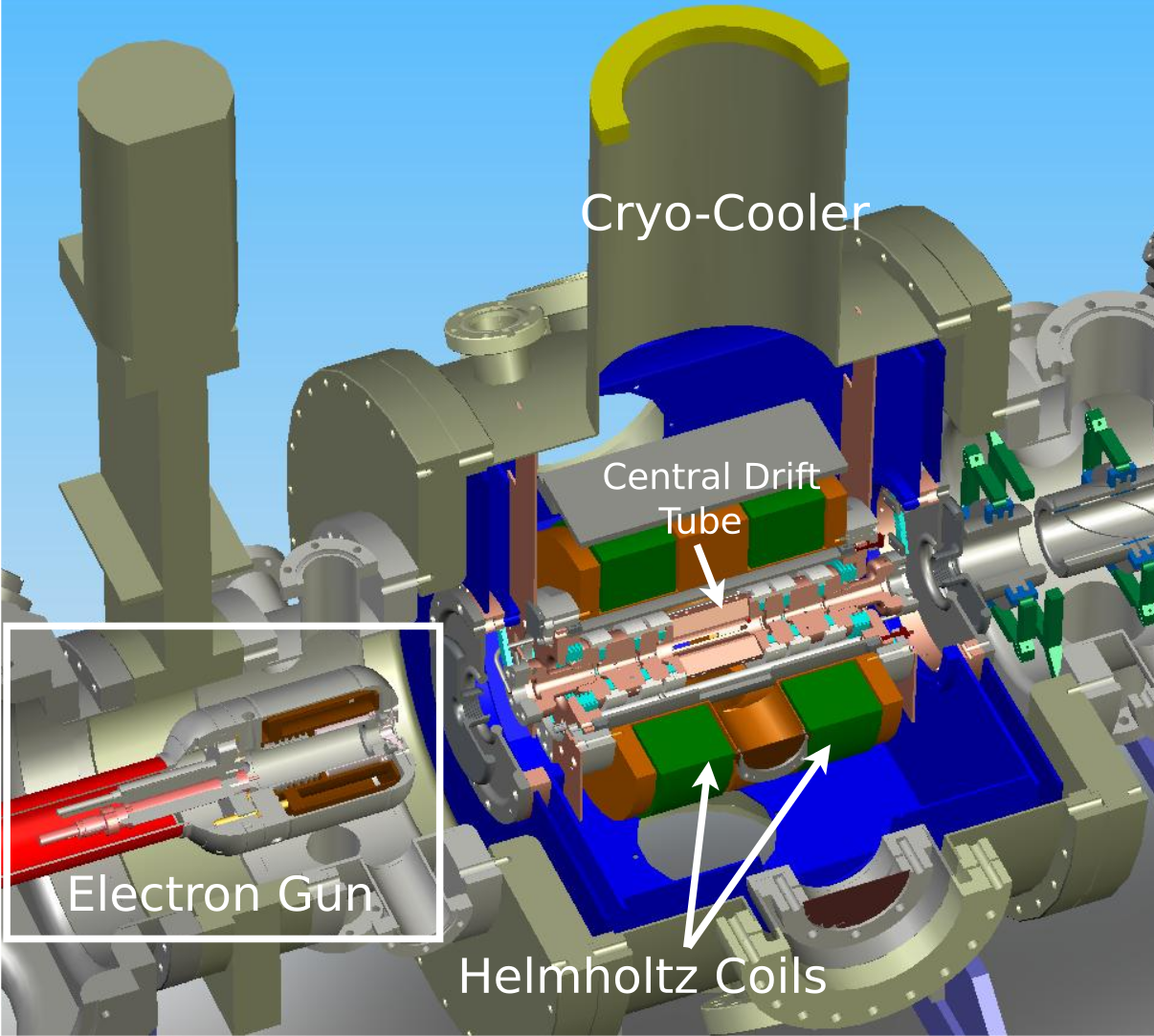}
\end{center}
\caption{\label{EBIT}(left) A photograph of the decay spectroscopy array in its configuration on the TITAN platform at TRIUMF.  Shown are the electron gun, two of the seven Si(Li) detectors, and the trap outline.  An empty support for one of the Si(Li) detectors is also shown, where the thin Be window that provides vacuum shielding for the trap can be seen at the centre. (right) A schematic view of the trap and electron-gun portions of the TITAN EBIT (adapted from Ref.~\cite{Lea14}).  The ion bunches enter and exit from the right of the image, and are subjected to ionization from the electron beam while they occupy the central drift tube.  The seven open access ports which surround the EBIT are occupied by planar Si(Li) detectors, which are used for the detection of low-energy X-rays, as discussed in the text.}
\end{figure}
The TITAN EBIT confines ions (i) axially by an electrostatic square-well potential, and (ii) radially by the electron-beam space-charge potential and magnetic field~\cite{Lap10}.  HCIs are generated by successive electron-impact ionization using a the 100~mA, 1.7~keV electron beam compressed by a strong magnetic field.  The up-to 6~Tesla magnetic field is produced by two superconducting Nb$_3$Sn coils in a Helmholtz-like configuration~\cite{Lap10}.  To observe the characteristic photons from the EC decays, each of the EBIT's seven external ports is occupied by a 5~mm thick Si(Li) crystal, which is optimized for the detection of X-rays~\cite{Lea14}.  A schematic view and an image of the TITAN decay-spectroscopy setup are displayed in Fig.~\ref{EBIT}.

\section{$^{116}$In Experiment}
\begin{figure}[t!]

\begin{center}
\includegraphics[width=\linewidth]{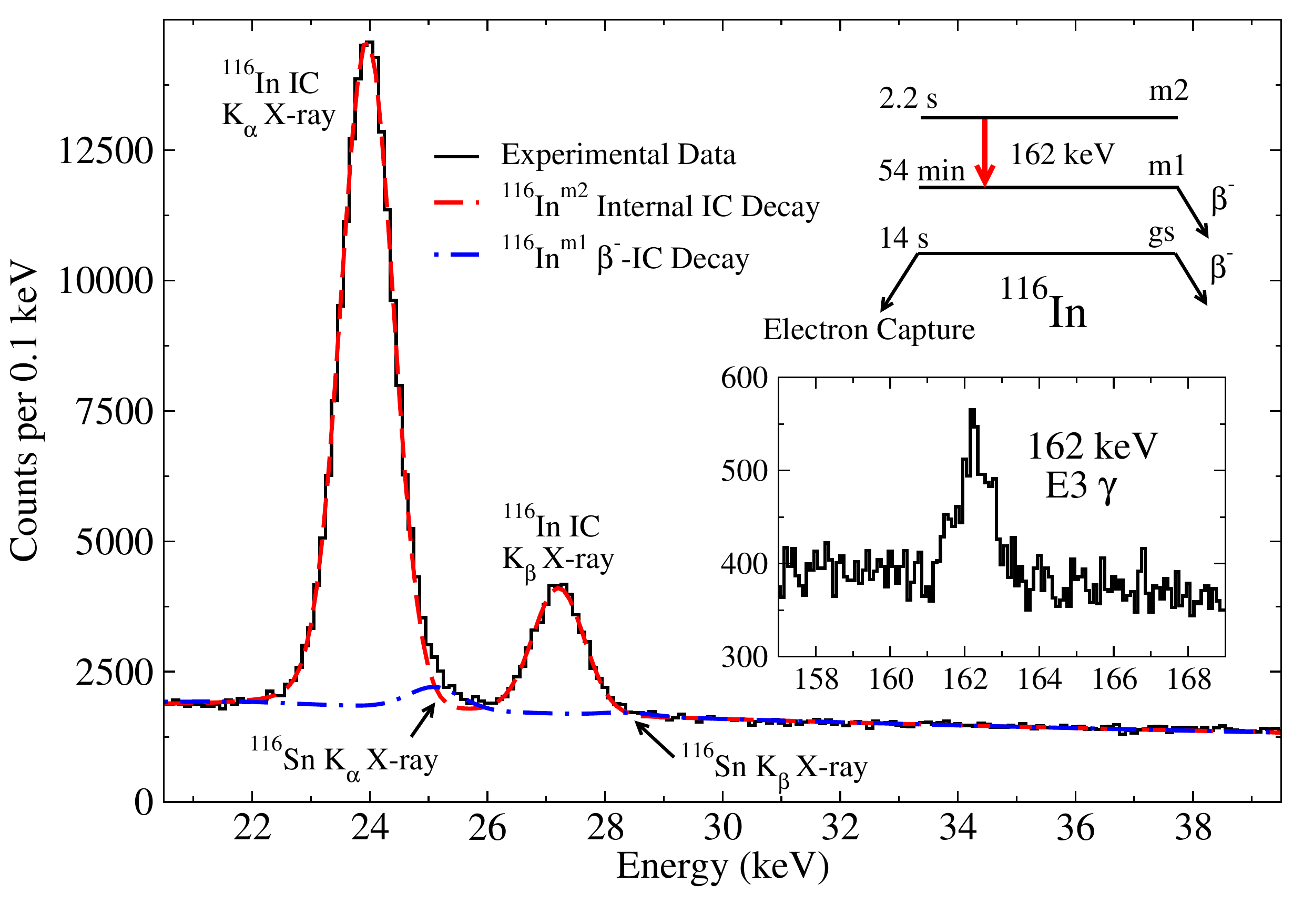}
\end{center}
\caption{\label{116InSpectrum}The detected x-ray spectrum from 20-40~keV for all runs taken during the experiment.  The dominant component in the observed decays resulted from the internal conversion of a 162~keV E3 transition between the $8^-$ $^{116}$In$^{m2}$ and the $5^+$ $^{116}$In$^{m1}$ states (shown in red), which has a characteristic half-life of 2.2~s.  The 162~keV E3 $\gamma$-ray from the same transition was also observed, however due to the low detection efficiency of the array at this energy, the observed statistics were low.  X-rays that resulted from internal conversion in $^{116}$Sn (blue) were also observed, however with a much lower intensity.}
\end{figure}
As a continued commissioning experiment, and a first attempt at one of the ECBR measurements relevant for $\beta\beta$-decay nuclear matrix elements, the $\beta$-decay of $^{116}$In was observed in the TITAN EBIT.  $^{116}$In is the intermediate nucleus in the $\beta\beta$ decay of $^{116}$Cd and decays primarily by $\beta^-$ with a half-life of 14.1~s~\cite{ENSDF}.  The 20~keV $^{116}$In ion beam from ISAC delivered to TITAN's RFQ consisted of roughly 10$^4$:10$^6$:10$^5$ ions/s of $^{116}$In$^{gs:m1:m2}$, respectively.  The ion bunches were subsequently injected into the EBIT (cycles are described below), and the summed photon spectrum that resulted from all 15~s decay cycles is displayed in Fig.~\ref{116InSpectrum}.  The dominant component in the spectrum is the K$_{\alpha}$ X-ray from the internal conversion of the 162~keV E3 transition from the 2.2~s $^{116}$In$^{m2}$.  Due to limited experimental time, and the large amount of isomeric contamination, the total collected statistics were insufficient to observe the weak ($\sim0.02\%$) electron-capture branch to $^{116}$Cd from the $^{116}$In$^{gs}$.

\subsection{Multiple Ion-Bunch Stacking}
\begin{figure}[t!]
\begin{center}
\includegraphics[width=0.8\linewidth]{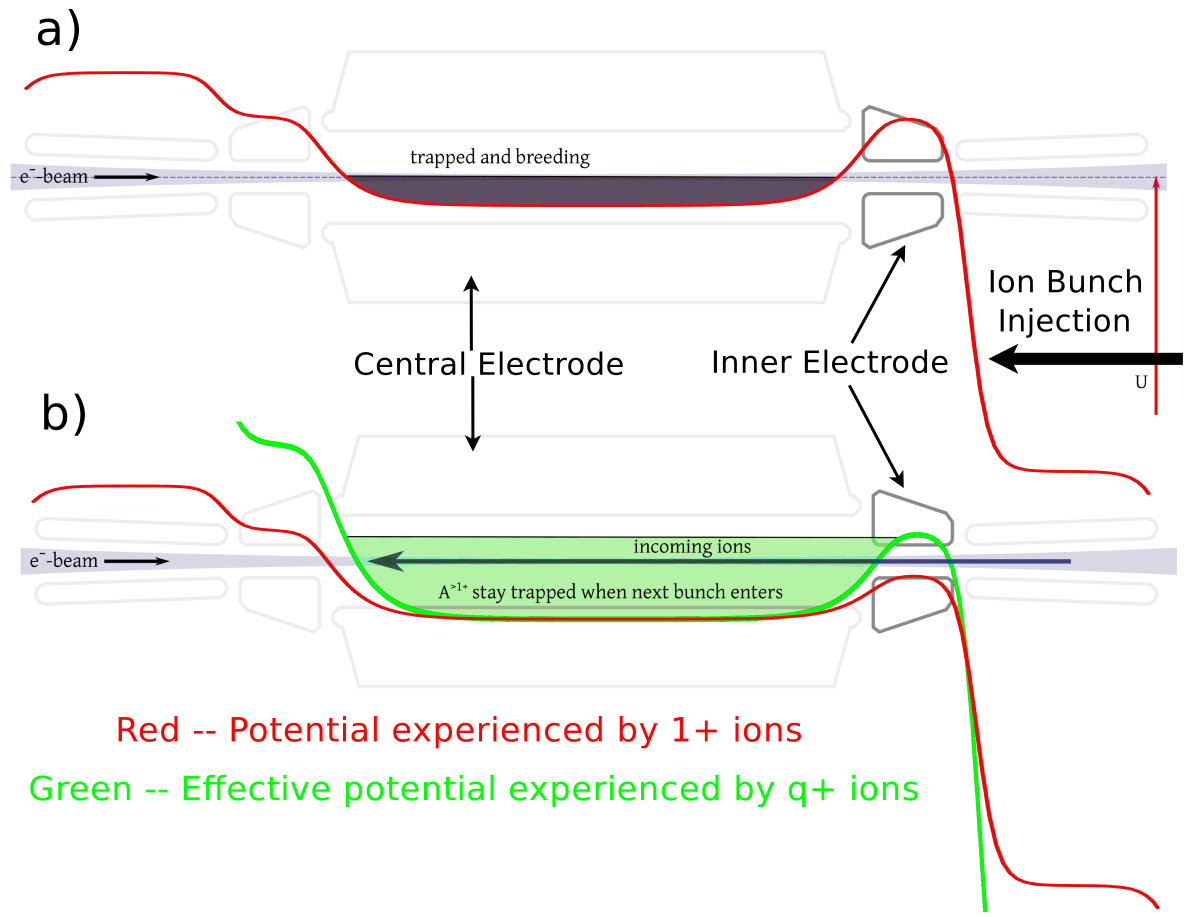}
\end{center}
\caption{\label{MIscheme}Schematic of the trapping potentials for (a) single injection and decay and (b) multiple injections before decay.  The single injection trapping scheme uses one cycle of filling the RFQ, injects the bunch into the EBIT, and closes the outermost trap electrode for storage and decay spectroscopy.  The multiple-injection scheme uses successive 25~ms extractions from the RFQ while the charge-bred ions in the EBIT experience a ``deeper" effective potential, and are not lost upon each injection.  After the space-charge limit of the EBIT is reached, the outermost electrode is raised for storage and decay spectroscopy.  This technique is described in further detail in the text.  Figure is adapted from Ref.~\cite{Kla14}.}
\end{figure}
Since the space-charge limit of the RFQ is $\sim10^5$-$10^6$~\cite{Lea14}, high ISAC beam intensities $\geq10^7$ for some species cannot be fully used and thus high production rates are wasted.  This is of particular concern for this experimental program, as the EC branching ratios relevant for the $\beta\beta$-decay cases~\cite{Fre07} are weak, ranging from $10^{-2}$ to $10^{-5}$, and require a large number of decays to observe them with any statistical significance.  A method for overcoming this space-charge limit was therefore tested using the $\beta^-$ decay of $^{116}$In by injecting many ion bunches into the EBIT without extraction~\cite{Ros13}.  To achieve this, the inner electrode potential is lowered for first ion-bunch injection and subsequently raised to confine the first ion bunch.  Following this, the injected ion bunch(es) quickly reach $q>2^+$ and remain confined during subsequent injections due to the increased effective potential experienced by the highly-charged ions.  The ions are then ejected, and the cycle is repeated.  This ion-bunch stacking technique is schematically displayed in Fig.~\ref{MIscheme}(b).

\begin{figure}[t!]
\begin{center}
\includegraphics[width=1.02\linewidth]{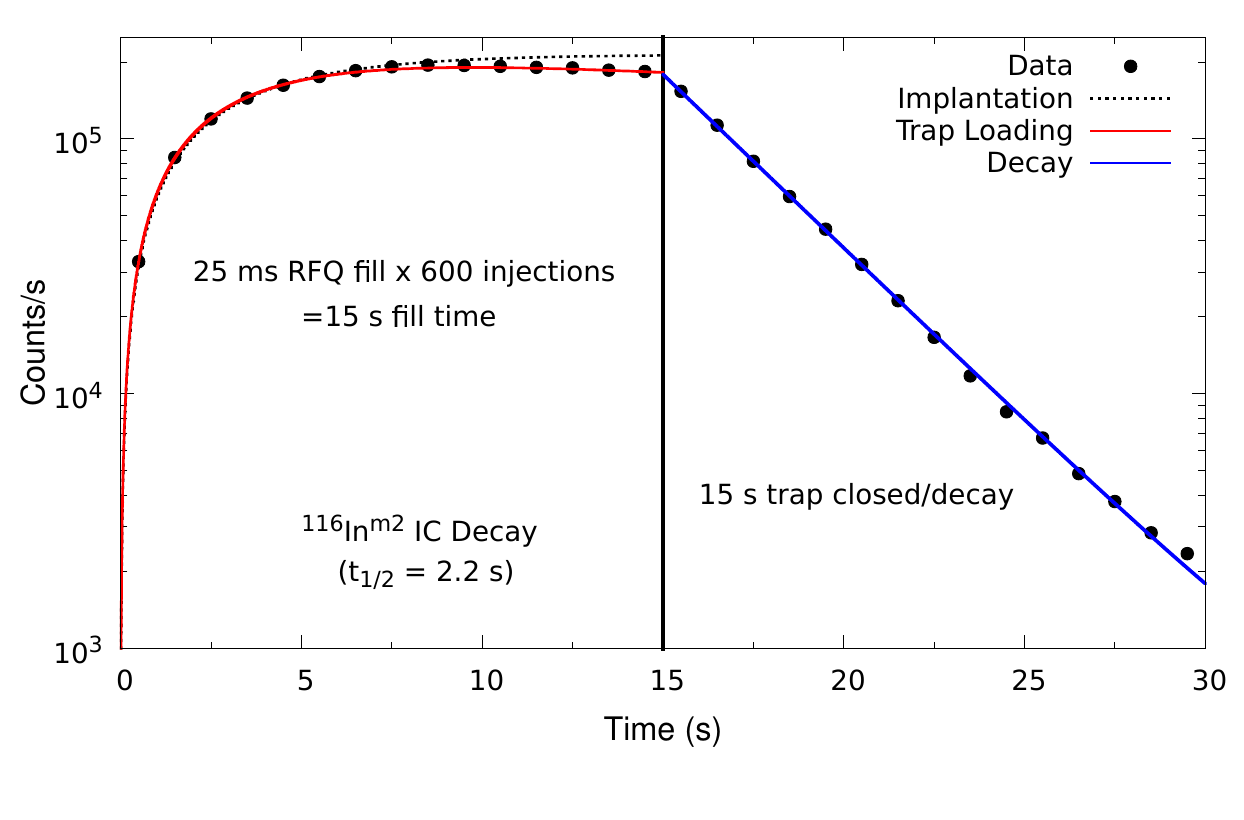}
\end{center}
\caption{\label{116InDecay}Time dependence of the number of counts observed in the K$_{\alpha}$ and K$_{\beta}$ X-rays resulting from the internal conversion from $^{116}$In$^{m2}$ ($t_{1/2}=2.2$~s).  The first 15~s illustrates the multiple-stacking technique discussed in the text, and shows a nearly $100\%$ trapping efficiency up to 9~s (relative to ion implantation), at which point the space-charge-limit of the trap is reached.  An analytic expression was developed to model the observed trap loading, shown in red, including the observed trend from 9-15~s.  The trap is subsequently closed at 15~s, at which point the data represents pure statistical radioactive decay of $^{116}$In$^{m2}$.  A fit to these data yields $t_{1/2}=2.2(1)$~s, which is in good agreement with Ref.~\cite{ENSDF} and demonstrates no additional trapping related losses which would manifest themselves as an additional component in the decay curve~\cite{Lea14}.  Error bars are present, but are smaller than the data points.}
\end{figure}

In order to investigate these new trapping effects, the cycles consisted of 15~s of bunch accumulation in the EBIT (from 600 ion-bunch injections at 25~ms RFQ fills), a 15~s decay measurement portion while the trap is closed, ion-bunch extraction from the trap, 5~s of background counting, and the cycles were repeated.  The observed injection and decay cycle is displayed in Fig.~\ref{116InDecay} for all data acquired during the experiment.  The trapping efficiency is nearly 100\% relative to ion-beam implantation $(1-e^{-\lambda t})$ until roughly 9~s, where saturation of the space-charge appears to limit injection of subsequent ion bunches.  Assuming a space-charge limit of the RFQ of $\sim10^5$, this cycle time corresponds to a maximum of roughly $10^7$ ions stored in the EBIT (at a charge-state of $q=25^+$).

\section{Conclusions}
In summary, an in-trap decay spectroscopy tool has been developed at TRIUMF using TITAN's electron-beam ion-trap.  The goal of this facility is to provide a low-background environment for the observation of weak EC branching ratios of the intermediate nuclei for $\beta\beta$ decay.  The ion-trap environment allows for the detection of low-energy photons by providing backing-free storage, while simultaneously guiding charged decay particles away from the trap center via the strong (up to 6~T) magnetic field.  The highly-charged ions also allow multiple ion-bunch stacking, whereby ion-bunches are injected into the EBIT in quick succession, without extraction, thus circumventing the space-charge limit of TITAN's RFQ.  The first demonstration of this technique with the EBIT was performed during the decay measurement of $^{116}$In, and displayed a near 100\% efficiency for trapping up to the $\sim10^9e$ space charge limit of the EBIT.  Loss mechanisms after this point cause a slight decrease in the overall injection performance, which are not yet fully understood.  The successful employment of the multiple injection technique opens the avenue for high-sensitivity experiments which were previously unfeasible due to statistical limitations.

\end{document}